\documentclass[useAMS,usenatbib]{mn2e}
\usepackage{epsfig}

\def\com{{\rm com}}
\def\dof{{\rm d.o.f}}

\def\grb{{\rm GRB}}
\def\hi{{\rm H\,{\sc i}}}

\def\iso{{\rm iso}}
\def\jet{{\rm jet}}
\def\lum{{\rm L}}
\def\m{{\rm m}}
\def\obs{{\rm obs}}

\def\rr{{\rm r}}
\def\sfr{{\rm SFR}}

\def\spi{{\em Spitzer}}
\def\swf{{\em Swift}}

\title[Star Formation History, GRBs and Metallicity] 
{Star Formation History up to \bf{\em{z}} = 7.4: Implications for Gamma-Ray Bursts
and the Cosmic Metallicity Evolution}
\author[Li-Xin Li]{Li-Xin Li\thanks{E-mail: lxl@mpa-garching.mpg.de}\\
Max-Planck-Institut f\"ur Astrophysik, 85741 Garching, Germany}

\begin{document}

%\date{\today}

\date{Accepted 2008 May 17. Received 2008 April 07; in original form 2007 October 16}

\pagerange{\pageref{firstpage}--\pageref{lastpage}} \pubyear{2006}

\maketitle

\label{firstpage}

\begin{abstract}
The current \swf\, sample of gamma-ray bursts (GRBs) with measured redshifts
allows to test the assumption that GRBs trace the star formation in the 
Universe. Some authors have claimed that the rate of GRBs increases with 
cosmic redshift faster than the star formation rate, whose cause is not known 
yet. In this paper, I investigate the possibility for interpreting the 
observed discrepancy between the GRB rate history and the star formation rate
history by the cosmic metallicity evolution, motivated by the observation that
the cosmic metallicity evolves with redshift and GRBs prefer to occur in low 
metallicity galaxies. First, I derive a star formation history up to redshift
$z=7.4$ from an updated sample of star formation rate densities obtained by 
adding the new UV measurements of Bouwens et al. and the new UV and infrared 
measurements of Reddy et al. to the existing sample compiled by Hopkins \& 
Beacom. Then, adopting a simple model for the relation between the GRB 
production and the cosmic metallicity history as proposed by Langer \& Norman,
I show that the observed redshift distribution of the \swf\ GRBs can be 
reproduced with a fairly good accuracy. Although the results are limited by 
the small size of the GRB sample and the poorly understood selection biases 
in detection and localization of GRBs and in redshift determination, they 
suggest that GRBs trace both the star formation and the metallicity evolution.
If the star formation history can be accurately measured with other 
approaches, which is presumably achievable in the near future, it will be 
possible to determine the cosmic metallicity evolution with the study on the 
redshift distribution of GRBs.
\end{abstract}


\section{Introduction}
\label{intro}

Since the discovery of the afterglows of gamma-ray bursts (GRBs) and the
determination of their redshifts \citep{met97,vanp97}, it has been firmly 
established that GRBs are at cosmological distances 
\citep[for recent reviews see][]{pir04,zha04,mes06}. Observations on 
the hosts of GRBs have revealed that long-duration GRBs (hereafter GRBs) are 
associated with faint, blue and often irregular galaxies with high star 
formation rates (SFRs) \citep[and references therein]{con05,fru06,tan07,wai07},
confirming the early speculation that GRBs occur in star-formation regions 
and arise from the death of massive stars (Paczy\'nski 1998; Wijers et al. 
1998; see, however, Le Floc'h et al. 2006). The discovery of the 
GRB-supernova connection \citep[and references therein]{gal98,li06,woo06b}
supports the collapsar model for long-duration GRBs (MacFadyen \& Woosley 
1999; MacFadyen, Woosley \& Heger 2001).

Because of their very high luminosity, GRBs can be detected out to the edge
of the visible Universe with minimal extinction by intervening gases and dust 
\citep{cia00,lam00,bro02,nao07} and are hence an ideal tool for probing the 
formation rate and the environments of stars at high redshift, the 
reionization history, as well as the cosmic chemical evolution 
\citep{fyn06b,pri06,sav06,tot06,bro07,cam07,gal07,pro07}. The advantage of 
GRBs over quasars for probing the high redshift Universe has been discussed 
by \citet{bro07}. It has also been proposed that GRBs can be used as standard 
candles to constrain cosmological parameters 
\citep[and references therein]{blo03,fri05,sch07}. However, this proposal has
been seriously challenged by a recent study of \citet{li07}.

\begin{figure}
%\vspace{2pt}
\includegraphics[angle=0,scale=0.51]{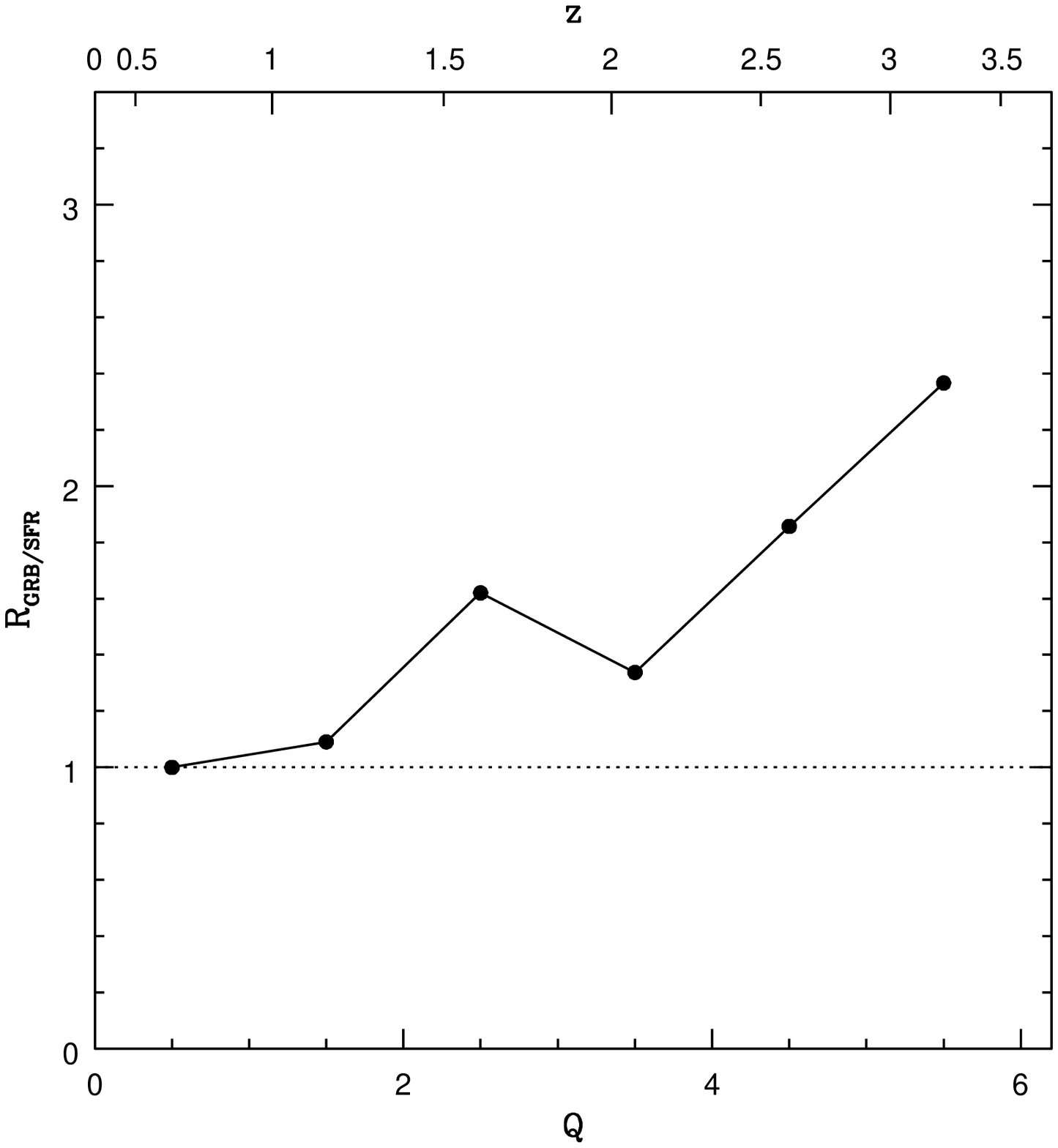}
\caption{The observed ratio of the GRB rate to the SFR, $R_{\grb/\sfr}$, as 
a function of $Q(z)$ and $z$, where $Q(z)$ is defined as an increasing 
function of redshift $z$ (equation \ref{Q} and Fig.~\ref{Q_z}). Normalization
is chosen so that $R_{\grb/\sfr} = 1$ at $Q=0.5$. The GRB rate history was 
obtained from a sample of {\em bright} \swf\ GRBs so that selection effects 
are minimized for $z\la 4$ \citep[see Section \ref{bright} of this 
paper]{kis07}. The SFR is the best fit to an updated sample of measured
cosmic SFRs up to $z\sim 7.4$ (eqs.~\ref{sfr} and \ref{ab}, Section 
\ref{sfh}). The horizontal dotted line denotes $R_{\grb/\sfr}=1$, an expected 
result of the assumption that GRBs trace the star formation unbiasedly. 
}
\label{rgrb_sfr}
\end{figure}

To study the relation between GRBs and the star formation, people often assume
that the GRB rate is proportional to the SFR then compare the predicted 
distribution of the GRB redshift (or other parameters, e.g. the intensity) to 
the observed distribution \citep[for a review see 
Coward 2007]{tot97,mao98,wij98,por01,nat05,jak06,dai07,le07}.
If GRBs trace the star formation in the Universe unbiasedly, one would expect 
that the ratio of the GRB rate to the SFR ($R_{\grb/\sfr}$) does not vary with
redshift. Then, an accurate measurement of the GRB rate history at high 
redshift would directly lead to the star formation history (SFH) in the early 
epoch which is otherwise hard to measure with the current technology because 
of the uncertainty in dust obscuration for UV photons \citep{hop06}. 

Unfortunately, recent studies show that GRBs do not seem to trace the star
formation unbiasedly (Fig.~\ref{rgrb_sfr}). Based on the current understanding
on the SFH and the \swf\ sample of GRBs with measured redshifts, people have 
found that $R_{\grb/\sfr}$ increases with redshift significantly 
\citep{dai07,le07,kis07,yuk07,cen08}. While observations consistently show that
the comoving rate density of star formation is nearly constant in the
interval $1\la z\la 4$ \citep{hop06}, the comoving rate density of GRBs 
appears evolving distinctly.

Adopting a model-independent approach by selecting bright \swf\, GRBs with
the isotropic-equivalent luminosity $L_\iso > 10^{51}$erg s$^{-1}$, 
\citet{kis07} found that there are $\sim 4$ times as many GRBs at redshift
$z\approx 4$ than expected from star formation measurements. They claimed
that some unknown mechanism is leading to an enhancement in the observed
rate of high-redshift GRBs. With a more sophisticated method, \citet{dai07} 
found that to reconcile the observed GRB redshift distribution with the 
measured SFH, the efficiency of GRB production by massive stars 
would be nearly six to seven times higher at $z\sim 7$ than at $z\sim 2$. 
Based on their results, Daigne et al. concluded that GRB properties or 
progenitors must evolve with cosmic redshift. 

In this paper, I investigate the relation between the SFH and the GRB rate 
history, and explore the possibility that the observed enhancement in the 
GRB rate at high redshift is caused by the cosmic metallicity evolution. 
Although the possibility of leading to an enhancement in the observed GRB 
rate evolution by the cosmic metallicity was mentioned by \citet{kis07}, they
did not give a quantitative analysis or a detailed discussion. Instead, 
Kistler et al. discussed more thoroughly on other possible causes, including
evolution in the fraction of binary systems which had been proposed as a 
channel for producing GRBs, an evolving initial mass function (IMF) of stars,
and evolution in the galaxy luminosity function (LF). 

There is growing evidence that metallicities play an important role in
the production of GRBs. Observations on the hosts of GRBs revealed that
GRBs prefer to occur in galaxies with low metallicities (Fynbo et al. 2003, 
2006a; Prochaska et al. 2004; Soderberg et al. 2004; Gorosabel et al. 2005; 
Berger et al. 2006; Savaglio 2006; Stanek et al. 2006; Wolf \& Podsiadlowski 
2007; Modjaz et al. 2008; Savaglio, Glazebrook \& Le Borgne 2008). Based on
the observational and theoretical evidence that the mass-loss rate of
Wolf-Rayet stars depends on the metallicity \citep{cro02,vin05,cro07},
theoretical studies on the collapsar model of GRBs arising from single massive
stars suggested that GRBs can only be produced by stars with metallicity 
$Z\la 0.1 Z_\odot$ since otherwise strong stellar winds will cause stars 
to lose too much mass and angular momentum to form a disk around a black hole
of several solar masses which is essential for the production of GRBs
(Hirschi, Meynet \& Maeder 2005; Yoon \& Langer 2005; Woosley \& Heger 2006;
Yoon, Langer \& Norman 2006).

It is well-known that the cosmic metallicity evolves strongly with redshift, 
and galaxies at higher redshift tend to have lower metallicities (Pettini et 
al. 1999; Prochaska et al. 2003; Rao et al. 2003; Kobulnicky \& Kewley 2004;
Kewley \& Kobulnicky 2005, 2007; Kulkarni et al. 2005, 2007; Savaglio et al.
2005, 2008; Wolfe, Gawiser \& Prochaska 2005; Savaglio 2006; P\'eroux et al. 
2007). \citet{nat05} considered a model where GRBs trace the average 
metallicity in the Universe rather than the SFR, and the GRB rate decreases 
with increasing metallicity. However, their model led to a GRB redshift 
distribution that is nearly indistinguishable from the distribution
predicted by the SFR \citep{jak06}. 

Recently, \citet{lan06} considered the effect of the cosmic metallicity 
evolution on the integrated production rate of GRBs in the framework of
the collapsar model, assuming that the GRB rate is jointly determined by the 
SFR and the metallicity evolution. Adopting a simple model for the metallicity
evolution and a best fitted SFR, and assuming that a GRB is produced if a 
progenitor star is massive enough and has a metallicity below a threshold 
($Z\la 0.1 Z_\odot$), they showed that the observed global ratio of the GRB 
rate to the core-collapse supernova rate ($\sim 0.001$) can be reproduced. 
\citet{nuz07} and \citet{cen08} investigated the host galaxies of GRBs in a 
cosmological hierarchical scenario with numerical simulations. They found that
the observed properties of GRB hosts are reproduced if GRBs are required to 
be generated by low metallicity stars. 

I will incorporate the model of \citet{lan06} into a probability distribution 
function of the luminosity and redshift of GRBs to study the rate history of
GRBs. For this purpose, I will present an updated SFH obtained by adding the 
new measurements of SFR densities at $z\sim 2.3$ and $3.05$ by \citet{red08}
and at $z\sim 3.8$, $5.0$, $5.9$ and $7.4$ by \citet{bou07,bou08} to the data
compiled by \citet{hop06}. With the updated data I will derive an analytic 
formula for the SFH and show that the observed distribution of \swf\ GRBs can
be successfully reproduced when the evolution of the cosmic metallicity is
properly taken into account. Then I will argue that, after a significantly 
expanded and well-defined sample of GRBs with measured redshifts and 
luminosities is available in future and the SFH at high redshift is accurately
determined with other approaches, GRBs will be a powerful tool for probing the
cosmic metallicity evolution.

The paper is organized as follows. In Section \ref{sfh}, I present the
updated measurements on the SFR up to $z\sim 7.4$ and derive an analytic 
formula for the cosmic SFH by fitting the updated data. In Section \ref{meta},
I summarize the current measurements on the evolution of the metallicity in
galaxies with cosmic time and argue that the data indicate a consistent
picture for the cosmic metallicity evolution when the redshift-dependent 
relation between the metallicity and the galaxy stellar mass is considered. 
In Section \ref{sample}, I describe the \swf\ GRB sample that is used for
the current work, show the luminosity distribution of the GRBs, and discuss
the selection biases involved in the detection and localization of \swf\ GRBs
and the measurement of their redshifts. In Section \ref{model}, the model 
that I adopt for calculating the GRB rate history is outlined, which includes 
assumptions about the probability distribution function, the SFR, the 
evolution of cosmic metallicity, and the form of the GRB LF. In Section 
\ref{results}, I present the results calculated with the model, and fit the 
observed distribution of the luminosity and redshift for the whole GRB sample 
and a bright GRB subsample. In Section \ref{conclusion}, I draw conclusions 
and discuss some implications of this work.

Throughout the paper, I assume a flat universe with $\Omega_\m = 0.3$,
$\Omega_\Lambda=0.7$, and $H_0 = 70$ km s$^{-1}$ Mpc$^{-1}$.

\section{The Updated Star Formation Rate History}
\label{sfh}

With modern UV and far-infrared (FIR) observations the SFH has been well 
established for redshift $z\la4$, with especially tight constraints for $z
\la 1$. \citet{hop06} compiled and critically analysed the data that were 
available then. They found that, despite large data scatter, the SFR in the 
redshift range of $1\la z\la 4$ is approximately a constant, which agreed with
the previous claim \citep[e.g.,][]{ste99}. Beyond $z\sim 4$, although the data
were highly incomplete, a meaningful constraint on the SFH was drawn: the SFR
declines with $z\ga 4$. From the 120 data points collected from UV, FIR, 
radio, H$\alpha$ and the
Hubble Ultra Deep Field (HUDF) estimates, \citet{hop06} critically selected 
56 `good' data points and fitted the SFH with simple analytical formulas and 
derived conservative uncertainties.\footnote{The number 58 of `good' data 
points printed in \citet{hop06} was a typo (A. M. Hopkins, private 
communications).}

For the criteria applied in the selection of `good' data and a complete
list of references for the data, please refer to \citet{hop06}. Here I update
the sample of \citet{hop06} by including the new measurements of
\citet{bou07,bou08} and \citet{red08}. 

Using the HUDF and the GOODS fields, \citet{bou07,bou08} found large samples
of star-forming galaxies at $z\sim4$, 5, 6, and 7--10. With those data, the 
rest-frame UV LFs were determined  with high accuracies at $z\sim 4$--6. It
has been found that the faint-end slope $\alpha$ and the normalization factor
$\phi^*$ show very little evolution with cosmic time, but the characteristic 
absolute magnitude $M_{\rm UV}^*$ brightens considerably from $z\sim 6$ to 
$\sim 4$ (by $\sim 0.7$ mag) \citep{bou07}. With all available deep optical 
and near-IR (NIR) data over the two GOODS fields, \citet{bou08} have also 
derived a rest-frame UV LF at $z\sim 7.4$, and obtained a constraint on the 
UV LF at $z\sim 10$. The SFR densities at $z\sim 4$--10 were derived from 
those UV LFs, confirming that the cosmic SFR density decreases quickly with
increasing redshift beyond $z\sim 4$.

Using a large sample of rest-frame UV-selected and spectroscopically observed 
galaxies in the redshift interval $1.9\le z\le 3.4$ combined with ground-based
spectroscopic H$\alpha$ and \spi\ MIPS 24 $\mu$m data, which includes over 
2000 spectroscopic redshifts and $\sim 15000$ photometric candidates in 29 
independent fields covering a total area of almost a square degree, 
\citet{red08} derived robust
measurements of the rest-frame UV, H$\alpha$, and IR LFs at $1.9\le z\le 3.4$.
The results indicate that the UV LF undergoes little evolution between $z\sim
4$ and $z\sim 2$. The SFR density at $z= 2.3\pm 0.4$ and $z=3.05\pm 0.35$ was
derived from the UV luminosity density and the IR luminosity density 
respectively, using the \citet{ken98} relations and assuming the \citet{sal55}
IMF from 0.1 to 100 $M_\odot$.

To include the new data points of \citet{bou07,bou08} and \citet{red08} into
the sample of \citet{hop06}, two types of corrections must be considered. 
First, UV lights are strongly obscured by dust so a dust-obscuration 
correction factor $C_1$ must be applied to the SFR density derived directly 
from the UV luminosity density. For $z\le 1$, the FIR SFR density was well 
measured with \spi\, \citep{flo05}. In \citet{hop06}, the UV data at $z\le 1$
were `obscuration corrected' by adding the FIR SFR density from \citet{flo05}
to each point. For obscuration corrections to the UV data between $z\approx 1$ 
and $z\approx 3$, \citet{hop06} made use of the
fact that the FIR measurements of \citet{per05} are quite flat in this domain 
as well as being highly consistent with those of \citet{flo05} at $z<1$, and
add the constant SFR density corresponding to that of \citet{flo05} at 
$z=1$. This looks a reasonable treatment for the obscuration correction at
$1<z<3$ since later measurements of the IR luminosity density 
\citep{cap07,red08} agree with a the trend found by \citet{per05} in this
redshift range \citep[fig.~27 of][]{red08}. In addition, the dust corrections
derived from the IR and UV SFRs of \citet{red08} are $C_1\approx 5.17$ at 
$z\sim 2.3$ and $C_1\approx 4.55$ at $z\sim 3$, in good agreement with
the values of $4.5$--$5.0$ at $z = 1.0$--$3.5$ obtained with other
approaches \citep{ste99,nan02,red04}.

\begin{figure}
%\vspace{2pt}
\includegraphics[angle=-90,scale=0.379]{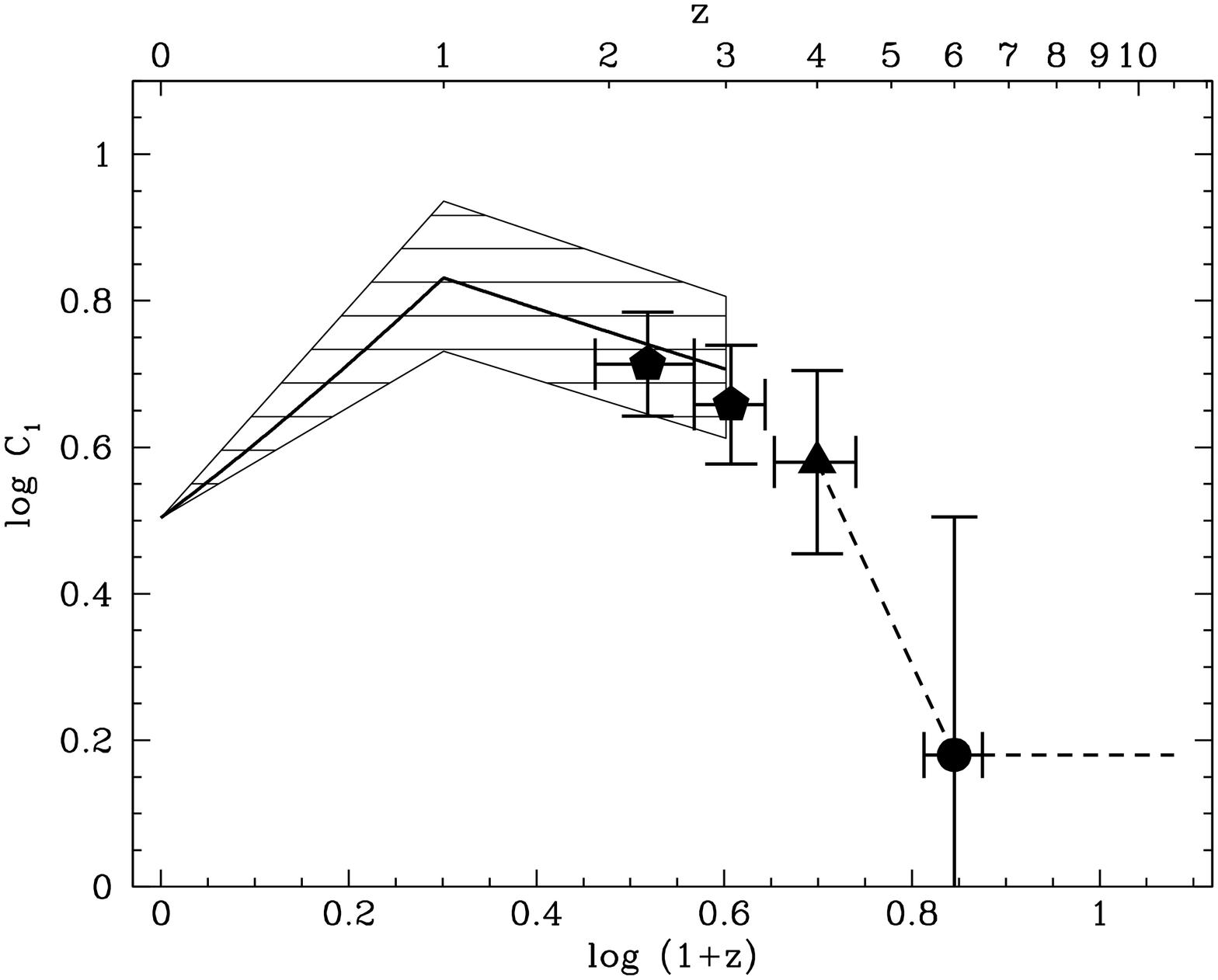}
\caption{Dust corrections for UV continuum emissions at different redshift 
intervals. The hatched region is the dust correction obtained from the FIR 
SFH of \citet{flo05} ($z<1$) and the FIR measurements of \citet{per05} 
($1<z<3$) (see text and Hopkins \& Beacom 2006 for details), relative to the 
UV SFH of \citet{sch05}. The two pentagons are estimated from the IR and UV 
SFRs of \citet{red08}, $C_1\approx 5.17$ at $z\sim 2.3$ and $C_1\approx 4.55$
at $z\sim 3$. The triangle is the dust correction $C_1 \approx 3.80$ at $z
\sim 4$ derived by \citet{ouc04}. The circle is the dust correction estimated
by \citet{bou06,bou07}, $C_1\approx 1.51$ at $z\sim6$. The dashed lines are
the linear interpolation between $z=4$ and $z=6$, and extension beyond $z=6$.
Note, the dust corrections beyond $z\sim 4$ are highly uncertain.
}
\label{sfr_dust_his}
\end{figure}

\begin{figure*}
%\vspace{2pt}
%\includegraphics[angle=-90,scale=0.373]{sfr_his.eps}
\includegraphics[angle=-90,scale=0.6]{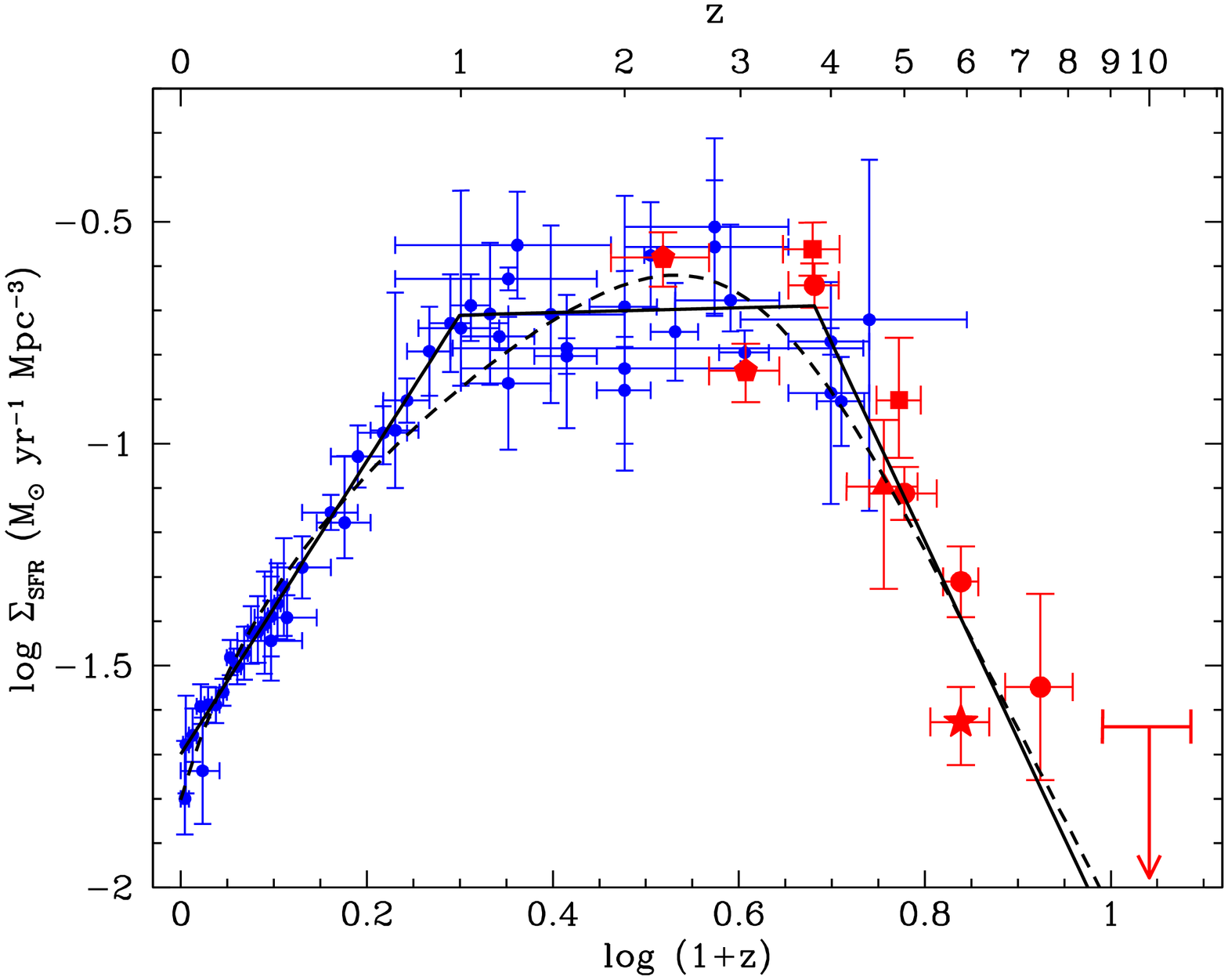}
\caption{Updated star formation history. The sample contains 51 data
points from \citet{hop06} (blue), two UV data points of \citet{gia04} with
updated corrections (red squares), one UV data point of \citet{bun04} with
updated corrections (red star), one UV data point of \citet{ouc04} with
updated corrections (red triangle), two UV+IR data points of \citet{red08}
(red pentagons), four UV data points and an upper limit of \citet{bou08}
(red circles and the downward arrow) (see text for details). The total number
of data points (not including the upper limit) is therefore 61. 
The solid black line is the best fit with a piecewise power-law
(eqs. \ref{sfr} and \ref{ab}), with $\chi^2_\rr = 1.18$ (55 degrees of 
freedom). The dashed black line is the best fit with a formula of \citet{col01}
(eq. \ref{cole}), with $\chi^2_\rr = 2.44$ (57 degrees of freedom).
In the calculation of chi-squares the errors in $z$ (bin sizes) are not taken
into account.
}
\label{sfr_his}
\end{figure*}

\begin{table*}
\centering
\begin{minipage}{170mm}
\caption{The star formation rate density derived from the new UV observational
data \citep{bou08} and from the old UV data with new corrections 
\citep{bun04,gia04,ouc04}
}
\label{sfr_data1}
\begin{tabular}{lllllll}
\hline
Reference & Estimator \hspace{0.3cm}~ & Redshift \hspace{0.7cm}~ & 
${\log \Sigma_{\sfr,{\rm un}}}^a$ \hspace{0.3cm}~ & ${\log \Sigma_\sfr}^b$ 
\hspace{0.8cm}~ & ${\log C_1}^c$ \hspace{0.cm} & ${\log C_2}^d$\\
\hline
\citet{gia04}......... & $\sim 1500$ \AA & $3.780\pm0.340$ & $-1.639\pm0.060$ & $-0.562\pm0.060$ & 0.580 & 0.496 \\
                       &         & $4.920\pm0.330$ & $-1.855\pm0.140$ & $-0.902\pm0.140$ & 0.379 & 0.575 \\
\citet{bun04}............. & $i$-Dropouts & $5.900\pm 0.500$ & $-2.301\pm0.079$ &
$-1.627\pm0.079$ & 0.180 & 0.494 \\
\citet{ouc04}............... & $\sim 1500$ \AA & $4.700\pm 0.500$ &
$-1.521\pm0.079$ & $-1.097\pm0.079$ & 0.424 & 0.000 \\
\citet{bou08}........... & $B$-Dropouts & $3.800\pm 0.300$ & $-1.720\pm0.050$ &
$-0.644\pm0.050$ & 0.580 & 0.496 \\
                       & $V$-Dropouts & $5.000\pm0.500$ & $-2.050\pm0.060$ &
$-1.112\pm0.060$ & 0.363 & 0.575 \\
                       & $i$-Dropouts & $5.900\pm0.300$ & $-2.180\pm0.080$ &
$-1.311\pm0.080$ & 0.180 & 0.689 \\
                       & $z$-Dropouts & $7.400\pm0.700$ & $-2.580\pm0.210$ &
$-1.548\pm0.210$ & 0.180 & 0.852 \\
                       & $J$-Dropouts & $10.00\pm1.200$ & $<-2.760$ &
$<-1.638$ & 0.180 & $0.942^\dagger$ \\
\hline
\end{tabular}\\
$^a$Uncorrected SFR density $\Sigma_{\sfr,{\rm un}}$ in units of 
$M_\odot\, {\rm yr}^{-1}\, {\rm Mpc}^{-3}$.\\
$^b$Dust and integration corrected SFR density, $\Sigma_{\sfr}= C_1 
C_2 \Sigma_{\sfr,{\rm un}}$, in units of $M_\odot\, {\rm yr}^{-1}\, 
{\rm Mpc}^{-3}$.\\
$^c$Dust correction factor $C_1$. At $z\sim 4$, I adopt $C_1 = 3.80$ 
\citep{ouc04}. At $z\ga 6$, I adopt $C_1 = 10^{0.18} = 1.51$ \citep{bou07}. 
For $4<z<6$, the value of $C_1$ is obtained by linear interpolation between 
$z=4$ and $z=6$ (Fig.~\ref{sfr_dust_his}). \\
$^d$Integral correction factor $C_2$, converting a `partial' SFR defined
by integration down to luminosity $L=L_{\min}$ to a `total' SFR defined by
integration down to $L=0$. The UV LFs published in \citet{bou08} are adopted,
with an average faint-end slope $\alpha=-1.71$. For the data of \citet{gia04}
and \citet{bou08}, $L_{\min} = 0.2 L^*_{z=3}$. For the data of \citet{bun04},
$L_{\min} = 0.1 L^*_{z=3}$. For the data of \citet{ouc04}, the 
$\Sigma_{\sfr,{\rm un}}$ has already been integrated down to $L=0$ by
\citet{hop06} so $C_2 =1$.\\
$^\dagger$$C_2$: For the upper limit at $z=10$ of \citet{bou08}, I take the
characteristic magnitude $M_{\rm UV}^*=-19.6$ in the calculation of $C_2$.
\end{minipage}
\end{table*}

Beyond redshift $z\approx 3$, dust obscuration is quite uncertain. 
Investigating the photometric properties of Lyman break galaxies (LGBs)
detected in Subaru Deep Field (SDF), \citet{ouc04} estimated the dust 
extinction of $z\simeq 4$ LGBs and obtained $E(B-V) = 0.15\pm0.03$. By the
\citet{cal00} extinction law, this redenning corresponds to a dust correction
factor $C_1\approx 3.80$. The dust extinction at $z\sim 6$ is still unknown.
Observing the fact that the UV continuium slope $\beta$ at $z\sim 6$ is bluer 
than that observed at $z\sim 3$ \citep{sta05,yan05,bou06}, \citet{bou06,bou07}
argued that the dust extinction at $z\sim 6$ is lower than that at $z\sim 3$. 
Adopting $\beta = -2$ \citep{bou06} and the extinction-$\beta$ relation 
$A_{1600}=4.43+1.99\beta$ \citep{meu99}, \citet{bou07,bou08} assumed that 
$C_1\approx 1.51$ (0.18 dex) at $z\sim 6$ in their SFR derivation.
However, comparison of the SFH derived from the Local Group and other nearby
galaxies with the cosmic SFH indicates a factor of $\sim 10$ extinction 
correction to high-redshift and UV-based SFR measures \citep{dro08}.

I will adopt $C_1\approx 3.80$ at $z\sim 4$, and $C_1\approx 1.51$ at $z\ga 
6$. At $4<z<6$, I will follow \citet{bou07,bou08} and estimate the dust 
correction by linear interpolation between the $z\sim 4$ and $z\sim 6$ 
values. The treatment for dust corrections outlined above is sketchedly
presented in Fig. \ref{sfr_dust_his}.

The next correction that should be considered is the factor that converts
a `partial' SFR calculated by integration down to a limit luminosity 
$L_{\min}>0$ to a `total' SFR calculated by integration down to $L=0$,
which is simply given by $C_2 = \Gamma(2+\alpha)/\Gamma(2+\alpha,L_{\min}/
L^*)$, where $\alpha$ is the faint-end slope of the Schechter LF, and $L^*$ 
is the characteristic luminosity. Here $\Gamma(x)$ is the gamma function, and
$\Gamma(a,x)$ is the incomplete gamma function.

The `partial' SFR calculated down to the survey magnitude limit is the SFR 
often reported in the literature. The `total' SFR is strongly denpendent on 
uncertainties in the faint-end slope, but it allows direct comparison with 
different SFR measurements \citep{sch05}. In the compilation of the SFR data 
in \citet{hop06}, it has been attempted to integrate to $L=0$ in all cases 
where possible, except for several data points which will be explained and
remedied below (A. M. Hopkins, private communications). To add the new data 
to the sample of \citet{hop06}, the integration correction must be made 
whenever it is relevant.

In the calculation of the integration correction factor $C_2$, I make use of 
the UV LFs derived by \citet{bou07,bou08} in the redshift interval $4\la z\la
10$. Since the fain-end slope $\alpha$ of the LFs shows very little evolution
with $z$, a mean value of $-1.71$ is used in all calculations.

The new data of \citet{bou08} and \citet{red08}, with the above corrections
being applied, are added to the `good' data compiled by \citet{hop06} and
shown in Fig. \ref{sfr_his}. Several data points of \citet{hop06} have been
modified or removed. Since \citet{bou06}'s SFR at $z\approx 6$ has been
updated to the new measurement by \citet{bou07,bou08}, it has been removed
from \citet{hop06}'s sample (then the sample contains 55 data points). The 
SFRs of \citet{gia04} at $z\approx 3.78$
and 4.92 and that of \citet{bun04} at $z\approx 5.9$, all in \citet{hop06}'s 
sample, were only integrated to a finite $L_{\min}$ ($= 0.2 L^*_{z=3}$ and
$0.1 L^*_{z=3}$, respectively; A. M. Hopkins, private communications). They
are updated by including the integration correction and with new dust
corrections according to the procedure described above. The SFR of 
\citet{ouc04} at $z\approx 4.7$ is updated with a new dust correction factor
obtained by linear interpolation between $z=4$ and $z=6$.

In Fig. \ref{sfr_his}, all new data (6 points, not counting the upper limit)
and updated old data (4 points) are shown with red symbols, and the unchanged
data in the sample of \citet{hop06} are shown with blue symbols. The new
data are broadly consistent with previous measurements, and extend the sample
to higher redshifts (up to $z\approx 7.4$). The new data and the updated
old data are listed in Tables \ref{sfr_data1} and \ref{sfr_data2}. Since
\citet{red08}'s SFRs are determined by combination of UV and IR measurements,
no extra dust correction is needed. The UV SFR is obtained by integration
down to $0.04 L^*_{z=3}$ which is already small enough. Since the UV SFR
contributes to the total UV+IR SFR only by a small fraction ($\sim 20\%$), a
further integration correction is not necessary. Hence the SFR data of
\citet{red08} are listed in a separated table. 

\begin{table}
%\centering
\caption{\citet{red08}'s star formation rate density derived from UV and IR
measurements (in $M_\odot\, {\rm yr}^{-1}\, {\rm Mpc}^{-3}$)
}
\label{sfr_data2}
\begin{tabular}{lllll}
\hline
Redshift\hspace{0.cm} & ${\log \Sigma_{\sfr,{\rm UV}}}^a$ &
${\log \Sigma_{\sfr,{\rm IR}}}^b$ & ${\log \Sigma_\sfr}^c$\\
\hline
$2.30\pm0.40$ & $-1.294_{-0.051}^{+0.045}$ & $-0.674_{-0.081}^{+0.068}$ & $-0.581_{-0.065}^{+0.057}$\\[1mm]
$3.05\pm0.35$ & $-1.494_{-0.070}^{+0.060}$ & $-0.944_{-0.090}^{+0.075}$ & $-0.836_{-0.071}^{+0.061}$\\
\hline
\end{tabular}\\
$^a$Uncorrected SFR density derived from the UV luminosity density, 
integrated down to $L = 0.04 L_{z=3}^*$.\\
$^b$SFR density derived from the IR luminosity density between 8 and 1000
$\mu$m.\\
$^c$The total SFR density obtained by the sum of $\Sigma_{\sfr,{\rm UV}}$ and 
$\Sigma_{\sfr,{\rm IR}}$.
\end{table}

In Fig.~\ref{sfr_his} and Tables \ref{sfr_data1} and \ref{sfr_data2} the
Salpeter IMF has been assumed. Conversion of SFH estimates to an alternative
IMF assumption corresponds to a simple scale factor, as described
by \citet{hop06}.

For the latter application, the SFH presented in Fig. \ref{sfr_his} is fitted
by a piecewise power-law, as was did in \citet{hop06}. The results are
\begin{eqnarray}
	\log \Sigma_{\sfr}(z) = a+b \log(1+z) \;,   \label{sfr}
\end{eqnarray}
where
\begin{eqnarray}
	(a,b) = \left\{\begin{array}{ll}
		(-1.70,3.30) \;, & z<0.993 \\
		(-0.727,0.0549) \;, & 0.993<z<3.80 \\
		(2.35,-4.46) \;, & z>3.80
		\end{array}\right. \;,
	\label{ab}
\end{eqnarray}
and $\Sigma_{\sfr}$ is in units of $M_\odot$ yr$^{-1}$ Mpc$^{-3}$. The
reduced chi-square of the fit is $\chi^2_\rr = 1.18$ (55 degrees of freedom;
errors in $z$ not counted). The value of $\chi^2_\rr$ indicates that the
fit is good.

The SFH is also fitted with a formula of \citet{col01}
\begin{eqnarray}
	\Sigma_{\sfr}(z) = \frac{a+b z}{1+(z/c)^d} \;. \label{cole}
\end{eqnarray}
The results are $(a,b,c,d) = (0.0157, 0.118, 3.23, 4.66)$, with $\chi^2_\rr 
= 2.44$ (57 degrees of freedom).

Compared to the fit results of \citet{hop06}, the updated sample leads to
a flatter SFH at $1\la z\la 4$, and a less rapidly decreasing SFH at $z\ga 4$.
A major reason for the large difference in the fitted high-$z$ slope 
($\sim -4$ versus $\sim -8$) is that in \citet{hop06} the SFR density of
Bunker et al. at $z=5.9$ has been underestimated by a factor $\sim 3$ since 
it has been integrated only down to $L_{\min}=0.1 L^*_{z=3}$. However, it 
should be stressed that at $z\ga 4$ the data suffer large uncertainties due 
to the lack of knowledge in dust obscuration as explained above.

\section{The Cosmic Metallicity}
\label{meta}

Observations have shown that the metallicity in the Universe evolves with 
cosmic time: the metallicity abundances in galaxies decrease with increasing
redshift. This is in agreement with the hierarchical scenario of structure
formation, in which metals are assumed to be produced by the formation of
stars, injected into the interstellar medium (ISM) by supernova explosions, 
and some of the metals are expelled from the host galaxy by supernova-driven
galactic winds. However, measurements of the cosmic metallicity evolution 
with different approaches do not appear to converge, and numerical 
simulations of galaxy formation are still not sufficient for fully 
reproducing the observations on the cosmic metallicity evolution (Nagamine, 
Springel \& Hernquist 2004; Kobayashi, Springel \& White 2007).

Metallicities in damped Ly$\alpha$ absorbers (DLAs) in the spectra of 
background QSOs have been extensively studied 
\citep{pet99,pro03,rao03,kul05,wol05}. The data consistently suggest that
the metallicity in QSO-DLAs evolves with cosmic time according to $Z/Z_\odot
\propto 10^{-\gamma z}$, with $\gamma \approx 0.2$--0.4 
\citep{pro03,nag04,kul05,kul07,sav06,per07}. In addition, there is evidence
indicating that the metallicity in sub-DLAs (absorbers with \hi\ column 
density $N_{\rm HI}<10^{20.3}$ cm$^{-2}$) evolves with redshift more strongly
than that in normal DLAs (with $N_{\rm HI}>10^{20.3}$ cm$^{-2}$), and are 
more metal-rich \citep{kul07,per07}.

DLAs have also been observed in the afterglows of GRBs, indicating 
metallicities and column densities that are generally higher than in QSO-DLAs
\citep{vre04,ber06,pro06,sav06,sud07,wat07,fyn08}. A system study on GRB-DLAs 
\citep{sav06} indicates that the metallicity in GRB-DLAs evolves with 
redshift with a slower rate ($\gamma\approx 0.18$) than that in QSO-DLAs, 
suggesting that GRB-DLAs and QSO-DLAs belong to two different classes of 
absorbers although they both trace the ISM in high-redshift galaxies.

By measuring nebular oxygen abundances for 204 emission-line galaxies with 
redshifts $0.3<z<1.0$ in the Great Observatories Origins Deep Survey-North 
(GOODS-N) field, \citet{kob04} found that the metallicity in galaxies with 
$-18.5<M_B<-21.5$ evolves with $\gamma\approx 0.14$ from $z=0$ to 1. 
Compilation of oxygen abundances in star-forming galaxies with $M_B<-20.5$ 
and $0<z<3.5$ indicates that the metallicity in star-forming galaxies
evolves with $\gamma\approx 0.15$ in the redshift range of $0<z\la 3$ 
\citep{kew05,kew07}.

Different types of metallicity measurements suffer different selection biases
so the fact that the values of $\gamma$ do not converge is probably not 
surprising. Detection of absorption lines in the neutral ISM of galaxies 
acrossing QSO sight-lines gives information of one line of sight in the 
galaxy so the result suffers large statistical fluctuations. Detecting 
emission lines in the integrated galaxy spectra is observationally challenging
at high redshift since prominent lines get very weak and redshifted to the 
NIR range \citep{sav05}. The different evolution observed for QSO-DLAs and
GRB-DLAs may indicate that GRB-DLAs are associated with more massive galaxies,
on average as massive as the LMC \citep{sav06}. Reconciling the metallicity 
distribution of GRB-DLAs, QSO-DLAs, and Lyman-break galaxies (star-forming 
galaxies) at high redshift has recently been discussed by \citet{fyn08}.

It is well-known that the stellar mass of star-forming galaxies is correlated
with the metallicity: galaxies with larger stellar masses tend to have higher
metallicities \citep[the mass-metallicity relation,][and references therein]
{sav05,erb06,kew08}. \citet{sav05} derived a simple relation that gives the 
metallicity in the gas of a galaxy at a given stellar mass and Hubble time 
(their eq. 11). With this relation, \citet{sav06} argued that the apparent 
difference in the values of $\gamma$ for QSO-DLAs and GRB-DLAs may be 
explained by the mass-metallicity relation with the assumption that different
measurements probe galaxies with different stellar masses.

Using the scaling $12+\log ({\rm O}/{\rm H}) = \log (Z/Z_\odot)+8.69$ 
\citep{sav06}, Savaglio's relation can be written as 
\begin{eqnarray}
	\log (Z/Z_\odot) &=& -16.28+2.53 \log M -0.0965 \log^2 M 
		\nonumber\\
		&&+5.17 \log t_{\rm H}-0.394 \log^2 t_{\rm H} \nonumber\\
		&&-0.403  \log t_{\rm H} \log M \;,
	\label{z}
\end{eqnarray}
where $M$ is the stellar mass in units of $M_\odot$, and $t_{\rm H}$ 
is the Hubble time in units of Gyr. 

The relation was used to fit the evolution of the metallicity in QSO-DLAs 
and GRB-DLAs. It was predicted that the stellar mass of QSO-DLAs is 
$10^{8.82\pm 0.65} M_\odot$ \citep{sav05}, and the stellar mass of GRB hosts 
is preferably in the range of $10^{8.6}$--$10^{9.8} M_\odot$ \citep{sav06}.
The latter is in agreement with the average stellar mass of $10^{9.3} M_\odot$
with a 1-$\sigma$ dispersion of 0.8 dex found for 46 GRB hosts with
optical-NIR photometry and spectroscopy \citep{sav08}.

With a galaxy stellar-mass function $\Phi(M)$, two average metallicities
can be defined. The first one is defined by averaging over the number of
galaxies
\begin{eqnarray}
	\left\langle\frac{Z}{Z_\odot}\right\rangle_1 \equiv \frac{
		\int_{M_{\min}}^\infty Z \Phi(M)dM}{Z_\odot 
		\int_{M_{\min}}^\infty\Phi(M)dM} \;.
	\label{Z_av1}
\end{eqnarray}
For a Schechter function of galaxy stellar-masses (Panter, Heavens \& Jimenez
2004), the low-mass-end slope $\alpha$ is usually $<-1$. Then, the integral 
of $\Phi(M)$ over $M$ does not converge as the lower limit of the integral 
goes to zero. Hence, a non-zero minimum stellar mass $M_{\min}$ is necessary
in the definition of $\langle Z/Z_\odot\rangle_1$.

The other average metallicity is defined by averaging over the stellar
mass (or the mass of \hi\ gases)
\begin{eqnarray}
	\left\langle\frac{Z}{Z_\odot}\right\rangle_2 \equiv \frac{
		\int_0^\infty ZM \Phi(M)dM}{Z_\odot \int_0^\infty M
		\Phi(M)dM} \;,
	\label{Z_av2}
\end{eqnarray}
i.e., the total metal mass divided by the total stellar (or \hi) mass.
In the definition of $\langle Z/Z_\odot\rangle_2$, the lower limit of mass
in the integral is set to zero since the integral converges.

\begin{figure}
%\vspace{2pt}
\includegraphics[angle=0,scale=0.511]{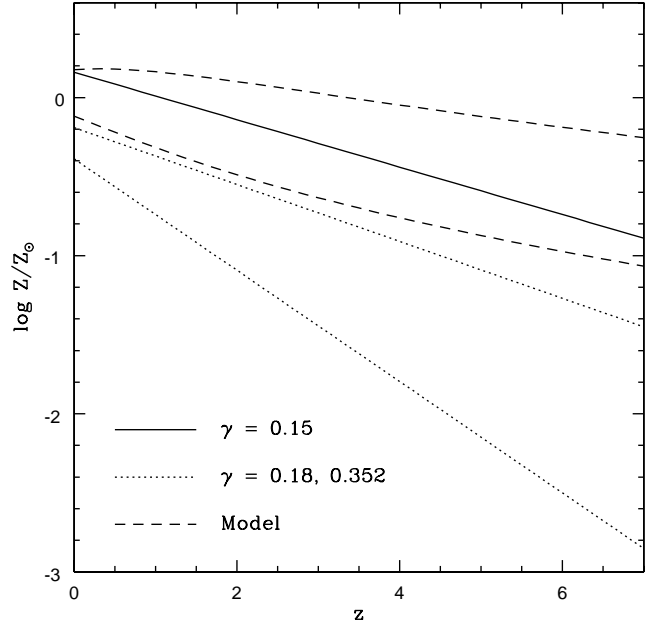}
\caption{The cosmic metallicity as a function of the redshift. The solid
line is the fit to the metallicity in star-forming galaxies by $Z/Z_\odot
\propto 10^{-\gamma z}$, with $\gamma = 0.15$ \citep{kew05,kew07}. The two
dotted lines are the fits to QSO-DLAs with $\gamma=0.352$ (lower line) and
GRB-DLAs with $\gamma=0.18$ \citep[upper line;][]{sav06}. The two dashed 
lines are the average metallicities calculated with the empirical model of 
\citet{sav05}, $\langle Z/Z_\odot\rangle_1$ (lower line, eq. \ref{Z_av1})
and $\langle Z/Z_\odot\rangle_2$ (upper line, eq. \ref{Z_av2}).
}
\label{Z_z}
\end{figure}

Adopting a Schechter function of galaxy stellar-masses from \citet{pan04}
with $\alpha = - 1.16$ and the critical mass $M_\star = 7.64\times 10^{10}
h^{-2} M_\odot$ (with $h=0.7$, $h$ is the Hubble constant in units of 100 
km s$^{-1}$ Mpc$^{-1}$), I calculated $\langle Z/Z_\odot\rangle_1$ (with 
$M_{\min} = 10^{7.5} M_\odot$) and $\langle Z/Z_\odot\rangle_2$ for the 
metallicity given by equation (\ref{z}). The results, with comparison to 
measurements, are shown in Fig.~\ref{Z_z}.

The $\langle Z/Z_\odot\rangle_1$ defined by equation (\ref{Z_av1}) is suitable
for comparison with observations since the observed metallicity evolution
is usually derived from the fit to the distribution of metallicity abundances
in individual galaxies. From Fig.~\ref{Z_z}, $\langle Z/Z_\odot\rangle_1$ (the
lower dashed line) is between the measured metallicity in star-forming
galaxies (with $\gamma=0.15$, solid line) and that in GRB-DLAs (with $\gamma
=0.18$, upper dotted line).

The $\langle Z/Z_\odot\rangle_2$ defined by equation (\ref{Z_av2}) (or its
variant, e.g., ratio of the total mass of metals to the total 
mass of \hi\ gases in all galaxies) is sometimes adopted in numerical 
simulations \citep[e.g.,][]{nag04}. From Fig.~\ref{Z_z}, $\langle Z/Z_\odot
\rangle_2$ (the upper dashed line) gives an estimate of the metallicity 
higher than that derived from the fit to measurements, and evolves with $z$ 
with a slower rate. This is caused by the fact that, by its definition, the 
contribution to $\langle Z/Z_\odot\rangle_2$ comes dominantly from galaxies 
with stellar masses around $M_\star\sim 10^{11} M_\odot$ while faster 
evolution and lower metallicities are attributed to galaxies with smaller 
stellar masses \citep{sav05}.

For $\alpha<-1$, the value of $\langle Z/Z_\odot\rangle_1$ and its evolution
are sensitive to $M_{\min}$, since the dominant contribution to the integral
comes from low stellar-mass and hence low metallicity galaxies because of 
their large numbers. The choice of $M_{\min}=10^{7.5} M_\odot$ seems 
reasonable according to the measurement on the distribution of stellar masses
for GRB host galaxies \citep{sav08}.

Although Savaglio's relation leads to reasonable results in fitting the 
metallicity distribution of QSO-DLAs and GRB-DLAs by varying the stellar 
mass, its validity and calibration need to be tested and improved with future 
measurements on metallicity and stellar masses of galaxies. For the
purpose of fitting the GRB rate history in this paper, it is enough to
adopt an empirical evolution equation
\begin{eqnarray}
	Z/Z_\odot\propto 10^{-\gamma z} \label{Z_evol}
\end{eqnarray}
with a constant $\gamma$ as adopted by \citet{lan06}. Given the limit number 
of GRBs with measured redshifts and spectra and many unknown observational 
biases, it is impossible to constrain the parameters in the metallicity 
evolution with GRBs at the present time. 

Since long-duration GRBs occurred in star-forming galaxies, following 
\citet{lan06} I will adopt the value $\gamma=0.15$ derived by 
\citet{kew05,kew07}.

\section{The {\em Swift} GRB Sample}
\label{sample}

\subsection{The Luminosity Distribution of {\em Swift} GRBs}
\label{luminosity}

Since the launch of \swf\ in late 2004, 310 GRBs have been detected by 31 
March 2008, of which 181 bursts have optical detections (by the UV/Optical
Telescope on \swf\ and/or ground telescopes) and 102 bursts have measured 
redshifts.\footnote{http://swift.gsfc.nasa.gov/docs/swift/archive/grb\_table/}$^{,}$\footnote{http://www.mpe.mpg.de/~jcg/grbgen.html}

\citet{but07} have compiled a catalog of 218 \swf\, GRBs and calculated their 
durations and spectral parameters between and including GRBs 041220 and 
070509, including 77 events with measured redshifts.
From that catalog, \citet{kis07} selected 63 bursts with long durations 
($T_{90} > 2$ s, $T_{90}$ is the observed time duration to contain 
$90\%$ of the total counts, with $5\%$ in front and the other $5\%$ behind.) 
and reliable redshift measurements to investigate the redshift distribution
of \swf\ GRBs. The following four long GRBs are 
contained in the catalog of Butler et al.  but not included in the sample 
of Kistler et al., because of their  unreliable redshifts: 060116, 060202, 
060428B and 061004 (M. D. Kistler, private communications).

\citet{lia07b} have derived the luminosity for 45 long \swf\ GRBs using the
method developed by \citet{zha07b}. \citet{cab07} have published the spectral 
and energetic results for a smaller sample of \swf\,GRBs, which contains 29 
long bursts with calculated isotropic-equivalent energy.

I use the sample of \citet{kis07} selected from \citet{but07}---with addition
of GRB 050223 ---for my analysis, since it is the largest sample to date. The 
64 GRBs in the sample have determined durations, redshifts, peak spectral 
energy, and the isotropic-equivalent energy in the 1--$10^4$ keV band in the 
rest frame of GRBs \citep{but07,koc07}. For the isotropic-equivalent energy 
$E_\iso$, I take the values from the table 1 of \citet{koc07} since they have 
more significant digits than the corresponding values in the table 2 of 
\citet{but07}. However, GRB 050223 (not included in Kistler et al. 2008) 
was not listed in \citet{koc07}, so I take the value of $E_\iso$ for this 
event from \citet{but07}.

\begin{figure}
%\vspace{2pt}
\includegraphics[angle=0,scale=0.47]{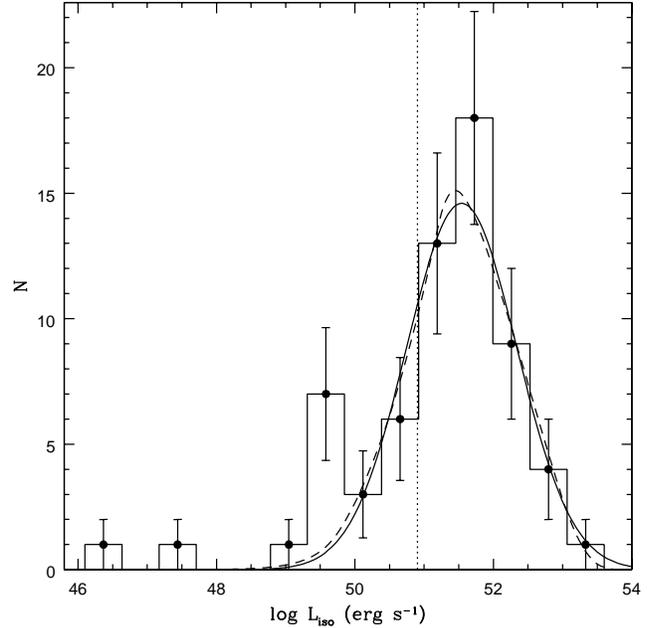}
\caption{Distribution of the isotropic-equivalent luminosity (defined by 
eq. \ref{liso}) for 64 long-duration \swf\, GRBs (the solid line histogram, 
with the number of GRBs in each bin indicated by a dark point with Poisson 
error bars). The solid curve is a Gaussian fit to the distribution of $\log 
L_\iso$, with $\chi_\rr^2 = 1.24$. The dashed curve is a least chi-squares 
fit by the model in Section \ref{model}, with $\chi_\rr^2 = 1.18$ (see 
Sections \ref{model} and \ref{results} for details). The 45 bursts to the 
right of the vertical dashed line at $E_\iso = 0.8\times 10^{51}$ erg 
s$^{-1}$ form a subsample of `bright GRBs', which are not affected by the 
luminosity threshold when $z<4$ (Section \ref{bright}).
}
\label{liso_distr}
\end{figure}

Following \citet{kis07}, I calculate the isotropic-equivalent luminosity of
a GRB by
\begin{eqnarray}
	L_\iso \equiv \frac{E_\iso}{T_{90}}(1+z) \;.   \label{liso}
\end{eqnarray}
The distribution of $L_\iso$ for the 64 GRBs in the sample is shown in 
Fig.~\ref{liso_distr}, which is fitted by a log-normal distribution with a 
mean of $\log L_\iso = 51.54$ ($L_\iso$ in erg s$^{-1}$), a dispersion 
$\sigma_{\log L_\iso} = 0.795$, and $\chi_\rr^2 = 1.24$.

The \swf\ trigger is very complex and the sensitivity of the detector is 
very difficult if not impossible to parameterize exactly \citep{ban06}. 
However, an effective luminosity threshold appears to be present in the data 
\citep[their figure 2]{kis07}. I find that the luminosity threshold can be 
approximated by a bolometric energy flux limit $F_{\lim} = 1.2 \times 
10^{-8}$erg cm$^{-2}$ s$^{-1}$. The luminosity threshold is then
\begin{eqnarray}
	L_{\lim} = 4\pi D_\lum^2 F_{\lim} \;,   \label{lum_lim}
\end{eqnarray}
where $D_\lum$ is the luminosity distance to the burst.

With the above luminosity threshold and an adopted GRB rate history, the
observed luminosity distribution can be fitted by an intrinsic Schechter
LF with a power-law index $-1.225$ and a characteristic luminosity $0.986
\times 10^{53}$ erg s$^{-1}$ (the dashed line in Fig.~\ref{liso_distr}; for 
details see Section \ref{all}).

The data have a moderate excess around $L_\iso =3.8\times 10^{49}$erg 
s$^{-1}$, at the $2.4$-$\sigma$ level (relative to the log-normal 
distribution). It probably indicates the existence of a faint population of 
GRBs, which will be discussed in details in Section \ref{results}.

\subsection{Selection Biases}
\label{biases}

Selection effects involved in a GRB sample are hard to model quantitatively.
There are at least two kinds of selection effects at work, which have
been extensively discussed in the literature 
\citep{blo03,cow07,fio07,gue07,kis07,le07,cow08}: 
(1) GRB detection and localization; and (2) redshift determination through 
spectroscopy/photometry of the GRB afterglow or the host galaxy. A full
address of these issues is beyond the scope of this paper, so I only give a 
brief description on them.

Since all the GRBs in the sample used in this paper \citep[and in][]{kis07}
are detected by Burst Alert Telescope (BAT) onboard \swf, biases arising from
GRB detection are minimized although the sensitivity of BAT is very difficult
to parameterize exactly \citep{ban06}. An effective luminosity threshold
introduced above and in \citet{kis07} would be a reasonable approximation of
the detection criteria for the bursts in the sample.

Due to the transient nature of GRBs, a fast and accurate location of the
burst is crucial for the determination of the redshift. The automatic slew
ability and the muti-wavelength nature of \swf\ have led that an accurate 
localization of the GRB is possible shortly after its detection, with an
efficiency higher than any previous GRB observatory (e.g., {\em BeppoSAX} 
and {\em HETE2}). An impressive number of ground-based facilities (dedicated
robotic telescopes, VLT, Gemini, Keck, etc) have been involved in \swf\
follow-up observations. However, optical afterglows have been discovered
for only $\sim 50\%$ of \swf\ GRBs, a fraction only slightly greater than
that of {\em BeppoSAX} and {\em HETE2} samples \citep{ber05,fio07,kan08}. 
The cause for the non-detection of optical afterglows for well localized 
GRBs remains a puzzle \citep{cas07,rol07,kan08}.

Generally, an apparently bright burst would be easier to localize than an
apparently faint burst. On average more distant bursts are expected to have 
lower observed gamma-ray fluxes, hence the number of GRBs with hight redshift
might have been somewhat underestimated. But this indicates that the deviation
of the GRB rate from that predicted by the SFR could be more serious than that
has been observed \citep{kis07}. 

For the GRBs detected and localized by \swf, most of them are followed-up
spectroscopically by ground-based telescopes. The participation of 
ground-based telescopes in follow-up observations has greatly increased the 
number of GRBs with redshifts (about $1/3$ of the total \swf\ GRBs). However, 
this also makes the selection biases in the redshift determination extremely 
complex, since instrumental selection biases of different telescopes have to
be involved. Different techniques have also been applied in the redshift 
measurement: absorption lines, emission lines and photometry of the Lyman 
edge, which causes additional biases. 

In addition to the various selection biases in redshift measurements discussed
above, there has been a well-known `redshift desert' in the redshift interval
$1.4\la z\la 2.5$ since galaxies in that range have been hard to detect
spectroscopically with traditional ground-based telescopes \citep{ade04,ste04}.
However, the `redshift desert' problem does not seem to be serious for \swf\
GRBs, since among the 85 long GRBs with measured redshifts by 31 March 2008,
23 bursts are in the redshift interval $1.4\la z\la 2.5$ 
(Fig.~\ref{qq_distr2u}). This is probably caused by the fact that different 
telescopes have different spectral ranges and they complement each other to 
some degrees.

\section{A Model for the GRB Rate History}
\label{model}

There is evidence that GRBs are beamed \citep{har99,kul99,sta99}. Hence,
when discussing the probability distribution function of GRBs, the effect
of jet beaming must be taken into account. Assuming that a GRB radiates its 
energy into two oppositely directed jets, each having a half-opening angle 
$\theta_\jet$. The total solid angle spanned by the jets is then $4\pi\omega$,
where the beaming factor $\omega\equiv 1-\cos\theta_\jet <1$.

For simplicity, I assume that except the comoving rate density, the property
of GRBs does not evolve with the cosmic redshift. Then, the intrinsic 
distribution function of $z$, $L_\iso$ and $y\equiv\log\left(\tan\theta_\jet
\right)$ must have a form 
\begin{eqnarray}
	P\left(z,L_\iso,y\right) = f(z) \phi \left(L_\iso\right) 
		\psi\left(L_\iso, y\right) \;,  \label{z_liso_y_dist} 
\end{eqnarray}
where
\begin{eqnarray}
	f(z) \equiv \left(\frac{c}{H_0}\right)^{-3}\frac{\Sigma_\grb(z)}
		{1+z}\frac{dV_\com}{d z} \;,
	\label{f_grb}
\end{eqnarray}
$(c/H_0)^{-3}\Sigma_\grb(z)$ is the comoving rate density of GRBs, and 
$V_\com$ is the comoving volume. Here $c$ is the speed of light.

In equation (\ref{z_liso_y_dist}), $\psi\left(L_\iso,y\right)$ is normalized 
with respect to $y$
\begin{eqnarray}
	\int_{-\infty}^{\infty} \psi\left(L_\iso, y\right) dy = 1 \;.
\end{eqnarray}

In a flat universe ($\Omega_\m+\Omega_\Lambda=1$), the comoving volume is
calculated by
\begin{eqnarray}
	\frac{dV_\com}{d z} = 4\pi D_\com^2 \frac{dD_\com}{d z} \;,
\end{eqnarray}
where the comoving distance
\begin{eqnarray}
	D_\com(z) \equiv \frac{c}{H_0} \int_0^z \frac{dz^\prime}{\sqrt{
	        \Omega_\m(1+z^\prime)^3 + \Omega_\Lambda}} \;.
\end{eqnarray}

A GRB is detected by an observer only if one of its jets points toward the 
observer. Hence, without consideration of other selection effects, the 
probability for a GRB to be detected by an observer is equal to $\omega$.
Multiplying equation (\ref{z_liso_y_dist}) by $\omega$ then integrating it
over $y$, one gets the observed distribution function of $z$ and $L_\iso$ 
without considering other selection effects
\begin{eqnarray}
	P\left(z,L_\iso\right) = f(z) \Phi \left(L_\iso\right) \;,  
	\label{z_liso_dist}
\end{eqnarray}
where the beaming-convolved LF
\begin{eqnarray}
	\Phi \left(L_\iso\right) \equiv \langle\omega\rangle\phi
		\left(L_\iso\right) \;, \hspace{0.5cm}
	\langle\omega\rangle \equiv \int_{-\infty}^{\infty}
		\omega \psi\left(L_\iso, y\right)\, dy \;.
\end{eqnarray}

Generally, the luminosity-averaged jet beaming factor $\langle\omega\rangle$
is a function of $L_\iso$. Therefore, the beaming-convolved LF $\Phi
\left(L_\iso\right)$ differs from the intrinsic LF $\phi\left(L_\iso\right)$.

\begin{figure}
%\vspace{2pt}
\includegraphics[angle=0,scale=0.46]{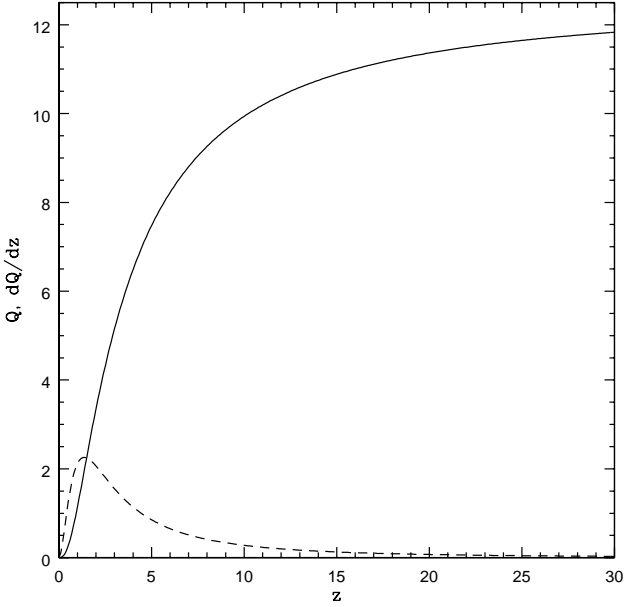}
\caption{The cosmic volume coordinate $Q$ (defined by eq. \ref{Q}) as a
function of the cosmic redshift $z$ (solid curve), and the derivative $dQ/dz$
(dashed curve). ($\Omega_\m = 0.3$, $\Omega_\Lambda=0.7$.)
}
\label{Q_z}
\end{figure} 

For the purpose of studying the rate density history of GRBs and the star
formation, it is more convenient to use a dimensionless volume coordinate
$Q$ than to use the redshift $z$, where $Q=Q(z)$ is defined by \citep{kis07}
\begin{eqnarray}
	Q(z) \equiv \left(\frac{c}{H_0}\right)^{-3} \int_0^z 
		\frac{1}{1+z^\prime}\frac{dV_\com}{d z^\prime} dz^\prime \;,
	\label{Q}
\end{eqnarray}
which increases monotonically with $z$ (Fig.~\ref{Q_z}). For $z\ll 1$, one has
$Q\approx 4\pi z^3/3$. As $z\rightarrow\infty$, one has $Q\rightarrow 
\mbox{constant}$.

The coordinate $Q$ is particularly useful in binning the data since the 
definition of $Q$ has taken into account both the effect of comoving volume
and the effect of cosmic time dilation. For example, if the comoving rate 
density of GRBs were a constant, in each equally-sized bin of $Q$ the number
of GRBs would be a constant. In contrast, if the data are binned with an equal 
size of $\Delta z$, the number of GRBs would change dramatically from bin to 
bin because of the comoving volume and the cosmic dilation factors in 
equation (\ref{f_grb}).

By equations (\ref{f_grb}) and (\ref{z_liso_dist}), the distribution
function of $Q$ and $L_\iso$ is
\begin{eqnarray}
	P(Q,L_\iso) = P(z,L_\iso)\frac{dz}{dQ} = \Sigma_\grb(Q)
		\Phi \left(L_\iso\right) \;.
	\label{Q_liso_dist}
\end{eqnarray}

Following \citet{lan06}, I assume that the GRB rate is related to the SFR and
the gas-phase metallicity in the host galaxy by 
\begin{eqnarray}
	\Sigma_\grb(z) = A \Psi(z,\epsilon) \Sigma_\sfr(z) \;,  
	\label{grb_rate}
\end{eqnarray}
where $\Sigma_\sfr(z)$ is the comoving rate density of star formation, 
$\Psi(z,\epsilon)$ is the fractional mass density belonging to metallicities 
below $Z = \epsilon Z_\odot$ at a given redshift $z$, and  $A$ is a
normalization factor. 

The parameter $\epsilon$ is determined by the metallicity threshold for the 
production of GRBs. Studies on the GRB progenitors and the collapsar model 
predict that long-duration GRBs are produced only for progenitor stars with 
$Z/Z_\odot\la 0.1$ \citep{hir05,yoo05,woo06b,yoo06,vin07}. Investigations on 
the host galaxies of long GRBs
show that an upper cut-off for the gas-phase metallicity of GRB hosts is 
likely $12+\log({\rm O}/{\rm H}) \sim 8.5$, corresponding to $Z\sim 
0.2$--$0.6 Z_\odot$ depending on the adopted metallicity scale and solar 
abundance vale (Modjaz et al. 2008; see also Wolf \& Podsiadlowski 2007, 
Savaglio et al. 2008). 

In general, the stellar metallicity is always lower than the gas-phase 
metallicity \citep{gal05}. Hence the two metallicity thresholds from the
theoretical study on GRB progenitors and the observational investigation on
GRB hosts appear to be consistent with each other. Since the metallicity
evolution adopted in equation \ref{grb_rate} and discussed in Section 
\ref{meta} refers to the gas-phase metallicity, throughout the paper 
I assume that $\epsilon = 0.3$.

According to \citet{lan06},
\begin{eqnarray}
	\Psi(z,\epsilon) = 1-\frac{\Gamma\left(\alpha+2,\epsilon^\beta
		10^{0.15\beta z}\right)}{\Gamma(\alpha+2)} \;,
	\label{psi_z_eps}
\end{eqnarray}
where $\alpha\approx -1.16$ is the power-law index in the Schechter 
distribution function of galaxy stellar masses \citep{pan04}, and $\beta
\approx 2$ is the slope in the linear bisector fit to the galaxy stellar 
mass-metallicity relation \citep{sav05,lan06}. In equation (\ref{psi_z_eps}),
it is assumed that the average cosmic metallicity evolves with redshift by 
$-0.15$ dex per unit redshift \citep[][see eq. \ref{Z_evol} in Section 
\ref{meta} and the relevant discussion]{kew05,kew07}.

For the SFR, \citet{kis07} adopted the piecewise power-law derived by 
\citet{hop06}. Here I use the the piecewise power-law derived from the 
updated data, given by equations (\ref{sfr}) and (\ref{ab}). The two formulas 
differ significantly only at $z\ga 4$, where measurements of the SFR are 
highly uncertain.

For the beaming-convolved LF of GRBs, I assume that it has a
Schechter-function form
\begin{eqnarray}
	\Phi \left(L_\iso\right) = \frac{1}{L_\star}\left(\frac{
		L_\iso}{L_\star}\right)^{\alpha_\lum}\exp\left(-L_\iso/
		L_\star\right) \;,
	\label{phi_liso}
\end{eqnarray}
where $\alpha_\lum$ and $L_\star$ are constant parameters to be determined 
by observational data. Although people often adopt a broken power-law or
a double power-law for the GRB LF \citep[e.g.,][]{gue05,lia07b}, I find that 
a simpler Schechter function is enough.\footnote{\citet{nat05} and 
\citet{sal07} also assumed a Schechter function for the GRB luminosity in 
their work.} Indeed, it appears that a power-law GRB LF at high end tends to
over-produce the number of bright GRBs (B. Zhang, private communications).

\section{Fitting the Redshift Distribution of {\em Swift} GRBs}
\label{results}

\subsection{Bright GRBs}
\label{bright}

To avoid the complication of a detailed treatment of the \swf\ detector's 
threshold and the assumption about the GRB LF, \citet{kis07} adopted a 
model-independent approach by selecting only GRBs with $L_\iso >10^{51}$erg 
s$^{-1}$ and $z<4$. The cut in luminosity and redshift minimizes the selection
effect in the GRB data. With 36 \swf\, GRBs that satisfy the above criteria, 
\citet{kis07} showed that the rate of GRBs increases with the redshift much 
faster than the SFR. A Kolmogorov-Smirnov test revealed that the SFR alone 
is inconsistent with the GRB rate history at the $95\%$ level.

For GRBs with $L_\iso>L_{\lim}$, the observed number of GRBs in an observer's
time interval $\Delta t_\obs$ and with $Q$ in the interval $Q$--$(Q+dQ)$ is 
$N(Q) \Delta t_\obs\, dQ$, where
\begin{eqnarray}
	N(Q) &\equiv& \int_{L_{\lim}}^\infty P(Q,L_\iso) dL_\iso \nonumber\\
		&=& \Sigma_\grb(Q)\int_{L_{\lim}}^\infty \Phi(L_\iso) 
		dL_\iso \;.
	\label{N0}
\end{eqnarray}
Submitting equation (\ref{phi_liso}) into equation (\ref{N0}), one gets
\begin{eqnarray}
	N(Q) = \Sigma_\grb(Q)\,\Gamma\left(1+\alpha_\lum,\frac{L_{\lim}}
		{L_\star}\right) \;.
	\label{N1}
\end{eqnarray}

There are 45 GRBs with $L_\iso > 10^{51}$erg s$^{-1}$ in the \swf\, sample, 
versus the 44 GRBs in \citet{kis07}. The difference in the two numbers is
caused by GRB 050318, whose luminosity is very close to $10^{51}$erg s$^{-1}$.
\citet{kis07} used the isotropic-equivalent energy of GRB 050318 published in 
\citet{but07} to calculate the luminosity and obtained a value that is slightly
below the luminosity cut. As mentioned in Section \ref{luminosity}, I take the 
isotropic-equivalent energy of GRBs from \citet{koc07}. I get a luminosity
for GRB 050318 that is slightly larger than $10^{51}$erg s$^{-1}$. To avoid 
this ambiguity arising from the luminosity uncertainty, hereafter I assume a 
luminosity cut $L_{\lim} = 0.8\times 10^{51}$erg s$^{-1}$. Then, the total 
number of GRBs with $L_\iso>L_{\lim}$ is unambiguously 45.

\begin{figure}
%\vspace{2pt}
\includegraphics[angle=0,scale=0.51]{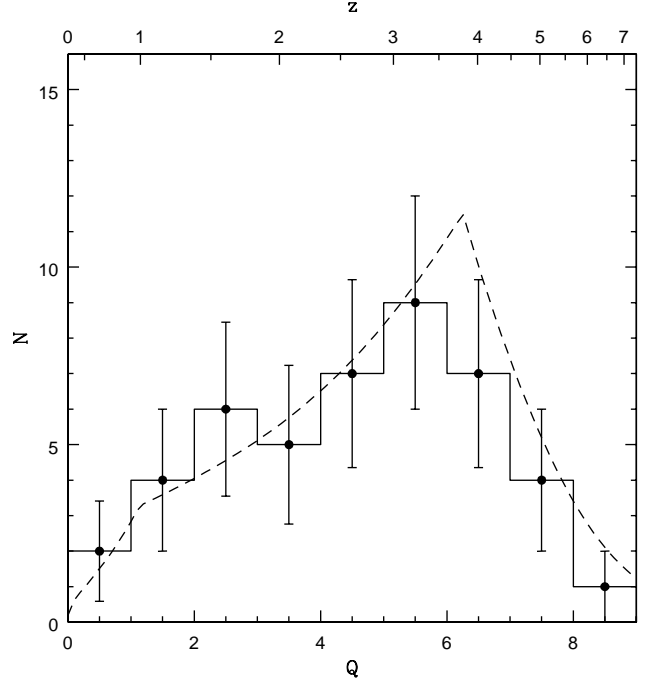}
\caption{Distribution of $Q$ for 45 \swf\, GRBs with $L_\iso>0.8\times 
10^{51}$erg s$^{-1}$ (the solid histogram, with the number of GRBs is each 
bin indicated by a dark point with Poisson error bars). The dashed curve is 
the best fit of the $\Sigma_\grb$ (eq. \ref{grb_rate} with $\Sigma_\sfr$
given by eqs. \ref{sfr} and \ref{ab}) to the 
first 6 data points ($z<4$) by varying the normalization, which has 
$\chi_\rr^2 = 0.14$. The deviation of the model from the data at $z>4$ is 
presumably caused by the flux limit of the detector which results a decrease 
in the number of detected GRBs.
}
\label{qq_distr}
\end{figure}

The distribution of $Q$ for the \swf\ 45 GRBs is plotted in 
Fig.~\ref{qq_distr}. The additional GRB 050318 falls in the third bin, 
resulting that the number of GRBs in the third bin is larger by one than 
that in \citet{kis07} (compare to their fig. 3).

Since $L_{\lim}$ is a constant, by equation (\ref{N0}) the number of GRBs
in each equally-sized bin of $Q$ is proportional to $\Sigma_\grb$, independent
of the GRB LF. For the selected bright GRBs, the luminosity 
threshold arising from the detector flux limit has no effect when $z\la 4$ 
\citep{kis07}. Using the $\Sigma_\grb$ given by equation (\ref{grb_rate}) with
$\Sigma_\sfr$ given by equations (\ref{sfr}) and (\ref{ab}) (and $\alpha=
-1.16$, $\beta=2$) to fit the first six data points in Fig.~\ref{qq_distr} 
which have $z<4$ by varying only the normalization, I get a surprisingly good
fit as shown by the dashed curve in the figure, with $\chi_\rr = 0.14$. This 
fact indicates that the GRB rate density assumed in equation (\ref{grb_rate}),
which takes into account the evolution of the cosmic metallicity, reasonably 
represents the true GRB rate. Hence, I will adopt this GRB rate density in 
all the following analysis.

\subsection{All GRBs}
\label{all}

Now I consider the effect of the detector flux limit (or, equivalently, the
luminosity threshold) on the distribution of 
luminosity and redshift for the 64 GRBs in the sample. As mentioned in 
Section \ref{luminosity}, the observed luminosity threshold can be modeled 
by an energy flux limit $F_{\lim} = 1.2 \times 10^{-8}$erg cm$^{-2}$ s$^{-1}$.
Then, $L_{\lim}$ is a function of $z$ (eq. \ref{lum_lim}). The observed 
distribution of $L_\iso$ is then given by
\begin{eqnarray}
	\hat{\Phi}\left(L_\iso\right) = \Phi\left(L_\iso\right)\Delta t_\obs
		\int_{0}^{z_{\max}} f(z) dz \;,
	\label{hat_phi}
\end{eqnarray}
where $z_{\max} = z_{\max}\left(L_\iso\right)$ is the maximum redshift up to
which a GRB of luminosity $L_\iso$ can be detected, solved from equation
$L_{\lim}(z)= L_\iso$.

Submitting equation (\ref{phi_liso}) into equation (\ref{hat_phi}) then
fitting the data in Fig.~\ref{liso_distr} with $\hat{\Phi}$, I get 
$\alpha_\lum = -1.225$, $L_\star = 0.986\times 10^{53}$erg s$^{-1}$, $A^\prime
\equiv\Delta t_\obs A= 9.386$, and $\chi_\rr^2 = 1.18$ (the dashed curve in 
Fig.~\ref{liso_distr}). The data excess around $L_\iso =3.8\times 
10^{49}$erg s$^{-1}$ is at the $2.3$-$\sigma$ level relative to the 
model.\footnote{I tried also to fit the data without the outlier point. But 
the main results in the paper are not affected drastically.}

The $N(Q)$ is still given by equation (\ref{N1}), but now $L_{\lim}$ is a 
function of $Q$. Since $-2<\alpha<-1$, the incomplete gamma function in 
equation (\ref{N1}) can be evaluated by the recurrence formula
\begin{eqnarray}
	\Gamma(a,x) = \frac{1}{a}\left[\Gamma(1+a,x)-x^a e^{-x}\right] \;.
\end{eqnarray}

\begin{figure}
%\vspace{2pt}
\includegraphics[angle=0,scale=0.51]{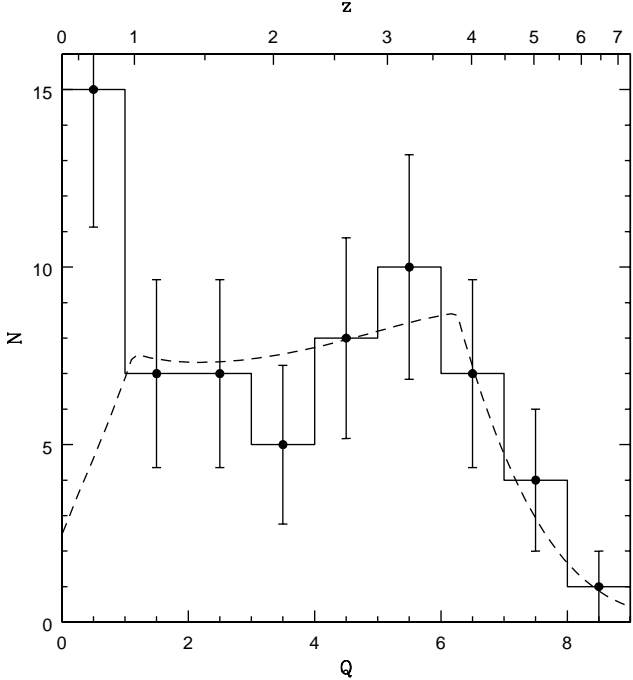}
\caption{Distribution of $Q$ for all the 64 \swf\, GRBs (the solid histogram,
with the number of GRBs in each bin indicated by a dark point with Poisson 
error bars). The dashed curve shows the distribution given by equation 
(\ref{N1}) with the parameters obtained by fitting the luminosity 
distribution (Fig.~\ref{liso_distr}, the dashed curve). The first data point 
has an offset about $2.7$-$\sigma$ from the dashed curve. The overall 
$\chi_\rr^2 = 1.01$. If the first data point is not included in the
calculation of chi-squares, then $\chi_\rr^2 = 0.24$. 
}
\label{qq_distr2}
\end{figure}

In Fig.~\ref{qq_distr2}, I show the distribution of $Q$ for all the 64 \swf\, 
GRBs in the sample. The dashed curve is the $N(Q)$ calculated by equation 
(\ref{N1}) with the normalization and the LF parameters 
determined above, and $L_{\lim}$ calculated by equation (\ref{lum_lim}). 
Globally, the modeled $N(Q)$ fits the observational data very well (without
adjustment of parameters), with $\chi_\rr = 1.01$. However, there is an 
obvious excess in the number of GRBs in the bin of $0<Q<1$, at the 
$2.7$-$\sigma$ level. If the first bin is excluded in calculating the 
chi-squares, I get $\chi_\rr = 0.24$.

The rate of bright GRBs in equation (\ref{N1}) with the normalization
obtained above and a constant $L_{\lim} = 0.8\times 10^{51}$erg s$^{-1}$ fits
the first six data points ($z<4$) in Fig.~\ref{qq_distr} with $\chi^2/\dof
= 0.682/6$, very close to the best fit obtained by varying the normalization
in Fig.~\ref{qq_distr} which has $\chi^2/\dof = 0.681/5$.

\subsection{The Cumulative Distribution of $Q$}
\label{cumulative}

\begin{figure}
%\vspace{2pt}
\includegraphics[angle=0,scale=0.497]{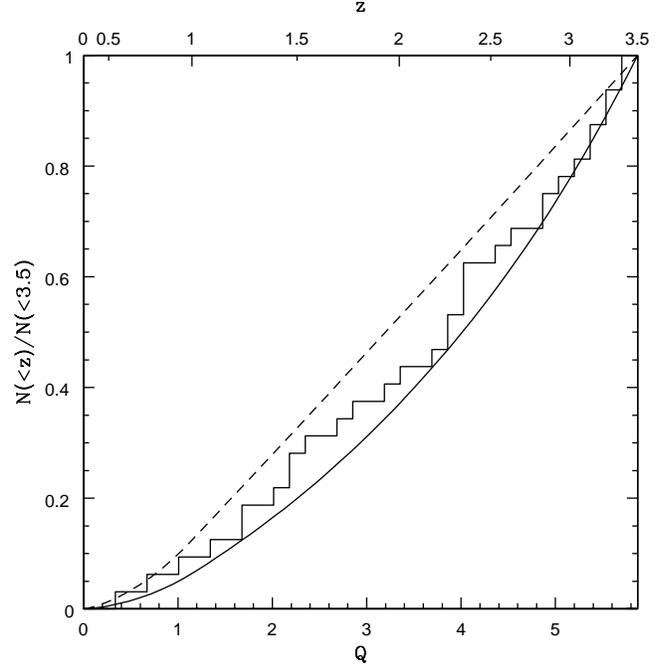}
\caption{Cumulative distribution of $Q(z)$ for 32 \swf\, GRBs with $z<3.5$ 
and $L_\iso>L_{\min} = 0.8\times 10^{51}$erg s$^{-1}$ (the stepwise curve). 
The cut in redshift and luminosity is chosen so that the data are not affected
strongly by the luminosity threshold. The smooth solid curve is calculated
with the GRB rate $\Sigma_\grb$ in equation (\ref{grb_rate}), which is 
independent of the GRB LF. The dashed curve is calculated with the SFR given 
by equations (\ref{sfr}) and (\ref{ab}) alone.
}
\label{qq_int}
\end{figure}

The cumulative distribution of Q for 32 \swf\, GRBs with $L_\iso >0.8 \times 
10^{51}$erg s$^{-1}$ and $z<3.5$ is shown in Fig.~\ref{qq_int}.\footnote{I set 
the upper bound of the redshift to $3.5$ instead of 4 to reduce the effect of 
the luminosity threshold.} This subsample of GRBs is not subject to the 
luminosity threshold. The integral distribution of $Q$, defined by
\begin{eqnarray}
	N(<z) \equiv \int_0^{Q(z)} N(Q) dQ  \;, \label{Nsz}
\end{eqnarray}
is shown with a smooth solid curve, which agrees with the observed data very 
well. The integral distribution of $Q$ given by the SFR alone (the dashed 
curve) does not agree with the observation.

\begin{figure}
%\vspace{2pt}
\includegraphics[angle=0,scale=0.51]{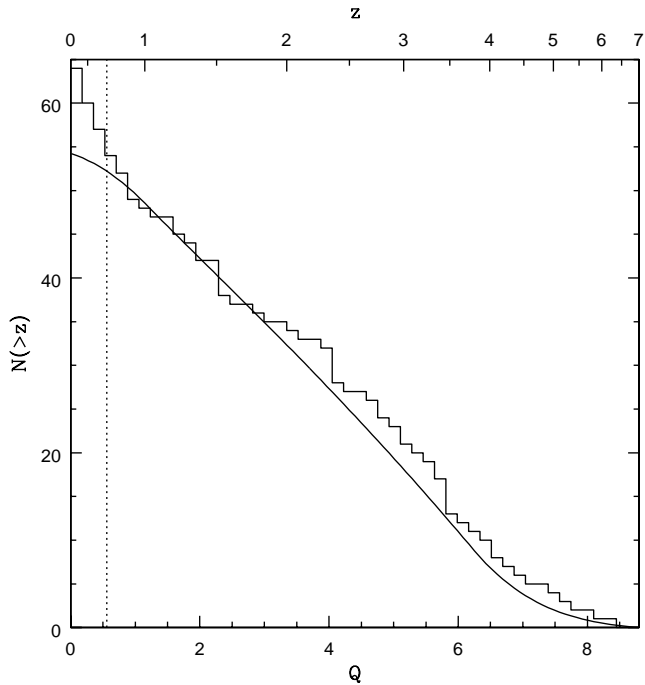}
\caption{Cumulative distribution of $Q(z)$ for all the 64 GRBs in the sample 
(the stepwise curve), with $N(>z)$ defined by equation (\ref{Ngz}) ($z_{\max}
= 7$). The smooth solid curve is the result given by the model. The vertical 
dotted line denotes $z = 0.7$ ($Q = 0.564$). An excess in the number of GRBs 
at $z< 0.7$ is clear (c.f. Fig.~\ref{qq_distr2}).
}
\label{qq_int2}
\end{figure}

In Fig.~\ref{qq_int2}, I show the cumulative distribution of Q for all the 64 
GRBs in the sample. To show the excess of GRBs at low redshift, I use the 
cumulative distribution defined by
\begin{eqnarray}
	N(>z) \equiv \int_{Q(z)}^{Q(z_{\max})} N(Q) dQ \;,  \label{Ngz}
\end{eqnarray}
where $z_{\max} = 7$. The $N(>z)$ given by the model is shown with a smooth
solid curve, which fits the observed distribution beyond $z=0.7$ very well. 
An excess in the number of GRBs at redshift $<0.7$ is clearly seen.

Given the small number of GRBs in the sample, the observed excess in the 
number of GRBs at low redshift (and low luminosity, Fig.~\ref{liso_distr}) 
could simply arise from statistical fluctuations. However, it is also possible
that the excess reflects a real deviation of the redshift (and luminosity) 
distribution from the model, because of the following facts:\\
(A) The detection of highly sub-luminous and sub-energetic nearby 
GRBs 980425, 031203 and 060218 has led people to propose that there exists 
a unique population of faint GRBs, whose rate is much higher than normal
cosmological GRBs \citep{cob06a,pia06,sod06,cha07,gue07,lia07b}.\\
(B) Some nearby long-duration GRBs are found not to have accompanied 
supernovae and hence are probably not related to the death of massive stars,
challenging the traditional scheme for the classification of GRBs by their 
durations \citep{geh06,zha06,zha07}. These non-supernova GRBs include 
060614 at $z=0.125$, 060505 at $z=0.089$ \citep{cob06b,del06,fyn06,geh06}, 
and 051109B at $z=0.08$ \citep{per06}.\\
(C) The SFH of the Local Group and other nearby galaxies indicates an excess
of the local SFR density relative to the cosmic SFH in the recent epoch of
$z\la 0.5$ \citep{dro08}. This might also be a cause for the excess in the
number of GRBs at $z\la 0.7$.

\section{Conclusions}
\label{conclusion}

I have presented an updated cosmic SFH up to redshift $z=7.4$. The updated
sample of SFR data are obtained by adding the new UV and IR measurements on
the SFR density of \citet{bou07,bou08} and \citet{red08} to the `good' data
sample compiled by \citet{hop06}. The two joined UV+IR measurements of
\citet{red08} at $z\sim 2.3$ and $z\sim 3.05$ agree well with previous
measurements of SFRs in the redshift interval $1\la z\la 4$, and are consistent
with a flat evolution SFH in this interval. The UV measurements of 
\citet{bou07,bou08} at $z\sim 3.8$, 5, 5.9, 7.4 and an estimate at $z\sim 10$
significantly expand the SFR data sample at $z\ga 4$, and are broadly 
consistent with the previous results in $3.5\la z\la 6$. The updated sample 
provides a consistent picture for the cosmic SFH up to $z\sim 7.4$
(Fig.~\ref{sfr_his}), although the dust correction at $z\ga 4$ is highly
uncertain which results large uncertainties in the SFH at high redshift
\citep{dro08}.

The updated SFH still under-produces the GRB rate density at high redshift 
when compared to the \swf\ GRB redshift distribution (Fig.~\ref{rgrb_sfr}),
confirming the previous claim \citep{dai07,le07,kis07}. The discrepancy is
investigated under the assumption that long-duration GRBs trace both the
star formation and the metallicity evolution, as motivated by the 
observations that long GRBs occurred in star-forming galaxies with low 
metallicities and the theoretical study on the collapsar model which shows that
GRBs can be produced only from low-metallicity massive stars.

Since the cosmic metallicity decreases with redshift 
\citep{pet99,pro03,rao03,kob04,kew05,kew07,kul05,sav05,sav06}, it is natural
to expect that the ratio of the GRB rate to the SFR must increase with 
redshift if the scenario that long GRBs are produced by the death of massive 
stars with low metallicity \citep{mac99,mac01,hir05,yoo05,woo06a} is correct.
Adopting a simple model for relating the GRB rate density to the SFR density
and the cosmic metallicity evolution \citep{lan06} and assuming a flux limit
for the \swf\ detector, I have shown that the redshift distribution of the 
64 \swf\ GRBs with measured redshifts and calculated luminosities can be 
successfully fitted by the updated SFH with a threshold in the metallicity 
for GRB production (Figs.~\ref{qq_distr}--\ref{qq_int2}).

\citet{kis07} have considered several possibilities for the cause of the
discrepancy between the the \swf\ GRB rate and the SFH. They have shown that 
the Kolmogorov-Smirnov test disfavors an interpretation as a statistical 
anomaly. Selection effects are also not likely to cause an increased 
efficiency in detecting hight-redshift GRBs. Although Kistler et al. have 
argued that alternative causes are possible (e.g., evolution in the fraction
of binary systems, an evolving IMF of stars, etc), the results in this paper 
indicate that the cosmic metallicity evolution may be the simplest 
interpretation.

\begin{figure}
%\vspace{2pt}
\includegraphics[angle=0,scale=0.51]{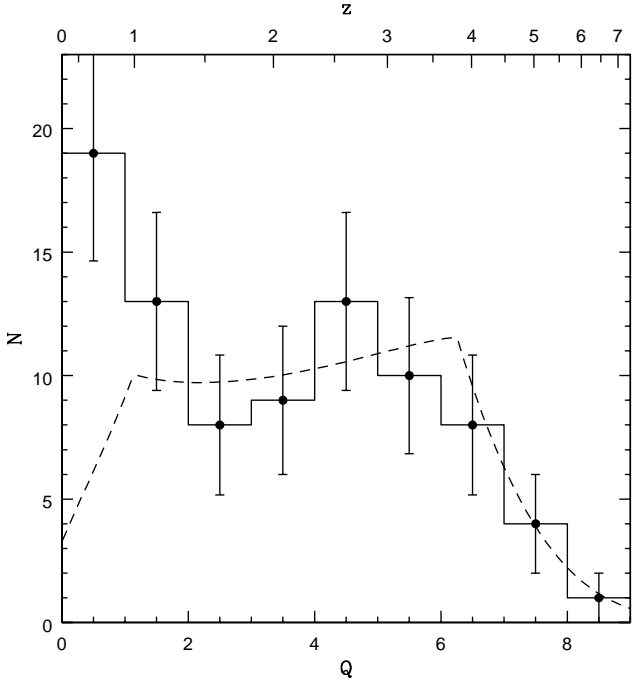}
\caption{Distribution of $Q$ for all the 85 \swf\, GRBs with redshifts, 
detected by 31 March 2008 between and including GRBs 041220 and 080330 
(the solid histogram). The dashed curve shows the distribution given by 
equation (\ref{N1}) with the parameters obtained by fitting the luminosity 
distribution (Fig.~\ref{liso_distr}, the dashed curve), renormalized by 
the total number of GRBs. The first data point has an offset about 
$3.0$-$\sigma$ from the dashed curve. The overall $\chi_\rr^2 = 1.21$. If
the first point is not included in the calculation of chi-squares, then 
$\chi_\rr^2 = 0.28$. 
}
\label{qq_distr2u}
\end{figure}

However, the results show an excess in the number of GRBs with low luminosity 
($L_\iso\sim 3.8\times 10^{49}$ erg s$^{-1}$; Fig.~\ref{liso_distr}) and at 
low redshifts ($z\la 0.7$; Figs.~\ref{qq_distr2} and \ref{qq_int2}). The 
existence of an excess is confirmed by the up-to-date \swf\ GRBs with measured
redshifts, detected by 31 March 2008 (Fig.~\ref{qq_distr2u}). Although it 
might simply be caused by statistical fluctuations, the observed excess could
also be consistent with the speculation that there is a unique population of 
intrinsically faint and nearby GRBs (Section \ref{results}).

The GRB sample used in this paper \citep[and in][]{kis07} is almost 
definitely incomplete and non-uniform, because of the complex selection biases
in the redshift measurement discussed in Section \ref{biases}. In fact, all 
the current works on the redshift distribution of GRBs have such a problem. 
However, \citet{kis07} have argued that none of the selection biases appears
to be able to increase the overall observability with redshift and account
for the enhancement in the GRB rate relative to the SFR.

The number of GRBs in the sample is small, which also prohibits one from 
obtaining a strict constraint on the parameters in the GRB LF and the cosmic 
metallicity evolution. 

Despite the above problems, the results of this work suggest that the rise of 
the observed \swf\ GRB rate relative to the SFR is compatible with an 
interpretation by the evolution of the cosmic metallicity. Once the problems
are solved or significantly alleviated in future, a significantly improved 
and enlarged sample of GRBs with measured spectra and redshifts will become 
available. Then, by comparing the observed GRB rate history to the SFH 
determined with other approaches, it will be possible to probe the cosmic 
metallicity evolution with GRBs.

\section*{Acknowledgments}

I thank R. J. Bouwens and A. M. Hopkins for helpful communications and 
sharing the data used in their papers, as well as M. D. Kistler, E.-W. Liang 
and B. Zhang for useful communications about their works. I also thank the 
anonymous referee for a very enlightening report.

\bsp

\label{lastpage}

\end{document}